\documentclass[conference,compsoc]{IEEEtran}
\IEEEoverridecommandlockouts
\usepackage{cite}
\usepackage{amsmath,amssymb,amsfonts}
\usepackage{algorithmic}
\usepackage{graphicx}
\usepackage{textcomp}
\usepackage{xcolor}
\usepackage{hyperref}
\usepackage[ruled,linesnumbered]{algorithm2e}

\begin{document}

\title{SMPC Task Decomposition: A Theory for Accelerating Secure Multi-party Computation Task}

\author{

\IEEEauthorblockN{
	Yuanqing Feng\IEEEauthorrefmark{1},
	Tao Bai\IEEEauthorrefmark{2},
	Songfeng Lu\IEEEauthorrefmark{3},
	Xueming Tang\IEEEauthorrefmark{3},
	Junjun Wu\IEEEauthorrefmark{3},
}

\IEEEauthorblockA{\IEEEauthorrefmark{0} Hubei Key Laboratory of Distributed System Security, Hubei Engineering Research Center on Big Data Security,\\ School of Cyber Science and Engineering, Huazhong University of Science and Technology,\\ Wuhan, 430074, China}
\IEEEauthorblockA{yuanqing\_feng@hust.edu.cn, d201980943@hust.edu.cn, lusongfeng@hust.edu.cn, xmtang@hust.edu.cn, junjunwu@hust.edu.cn}
}

\maketitle

\begin{abstract}
Today, we are in the era of big data, and data are becoming more and more important, especially private data. Secure Multi-party Computation (SMPC) technology enables parties to perform computing tasks without revealing original data. However, the underlying implementation of SMPC is too heavy, such as garbled circuit (GC) and oblivious transfer(OT). Every time a piece of data is added, the resources consumed by GC and OT will increase a lot. Therefore, it is unacceptable to process large-scale data in a single SMPC task.

In this work, we propose a novel theory called SMPC Task Decomposition (SMPCTD), which can securely decompose a single SMPC task into multiple SMPC sub-tasks and multiple local tasks without leaking the original data. After decomposition, the computing time, memory and communication consumption drop sharply. We then decompose three machine learning (ML) SMPC tasks using our theory and implement them based on a hybrid protocol framework called ABY. Furthermore, we use incremental computation technique to expand the amount of data involved in these three SMPC tasks. The experimental results show that after decomposing these three SMPC tasks, the time, memory and communication consumption are not only greatly reduced, but also stabilized within a certain range.
\end{abstract}


\section{Introduction}
Machine learning (ML) is a branch of artificial intelligence (AI) that aims to build systems that learn from data to predict future events or classify the unknowns. ML techniques are widely used in the business field, financial field and medical field. For example, systems built using ML can recommend products to users, analyze the rise and fall of stocks, and diagnose patients' conditions. In addition, ML techniques can also be used in image recognition, speech recognition, traffic forecasting, autonomous driving, etc.

Training an accurate predictive model usually requires a large amount of data. However, collecting data is difficult, especially private data. SMPC enables different organizations to jointly train a model without revealing their private data. Many SMPC-based ML algorithms have been implemented, such as neural networks \cite{9219246}, linear regression\cite{article}, logistic regression \cite{7958569}, support vector machines \cite{Privacy-preservingSVMclassification}, decision trees \cite{10.1145/335191.335438}, principal component analysis \cite{8073817} and k-mean clustering \cite{10.1145/1081870.1081942} \cite{10.1145/1315245.1315306} \cite{Mohassel2019PracticalPK}, etc. However, as the amount of data increases, the time, memory, and communication consumed to train a model also increase linearly. Therefore, it is difficult to train a model through one SMPC task when there are too much data involved in training.

\subsection{Related Work}
Many state-of-the-art techniques have been proposed to improve the performance of SMPC. We group these techniques into the following categories:
\begin{enumerate}[\IEEEsetlabelwidth{12}]
	\item 
	\textbf{System and Framework.} ABY3 \cite{Mohassel2018ABY3AM} is a framework that allows three parties to participate in SMPC task calculations. The CRYPTEN \cite{Knott2021CrypTenSM} is a software framework to foster adoption of SMPC. CryptGPU \cite{9519386} is a system for privacy-preserving machine learning that implements all operations on the GPU. Aarushi Goel \cite{10.1007/978-3-031-06944-4_14} provided a generic framework for branching multi-party computation that supports any number of parties.
	\item [2)] 
	\textbf{Protocol}. S. Dov Gordon \cite{10.1007/978-3-030-77886-6_24} proposed a suite of novel MPC protocols that maximize throughput when run with large numbers of parties. David Heath \cite{10.1007/978-3-030-56880-1_27} proposed Stack that eliminates the communication cost of inactive conditional branches. 
	\item [3)] 
	\textbf{Library}. SecFloat \cite{9833697} is a library for secure 2-party computation of 32-bit single-precision floating-point operations and math functions. SiRnn \cite{Rathee2021SiRnnAM} is a library for end-to-end secure 2-party DNN inference.
\end{enumerate}

\subsection{Motivation}
While the above techniques can improve the performance of SMPC, it is still unrealistic to train models on large-scale data through one SMPC task. Currently, it is difficult to significantly improve the performance of SMPC, so we try to find a new way to speed up SMPC tasks. We find that the SMPC task of combining data from different parties using operations satisfying commutativity and associativity can be decomposed. Because operations satisfying commutative and associative properties are decomposable. Parties can decompose one SMPC task into multiple local tasks and multiple SMPC sub-tasks. Then, parties can use one SMPC sub-task to combine the results of one local task. This method greatly speeds up calculations and reduces resources consumption.

\subsection{Our Contributions}
Our contributions are as follows:
\begin{enumerate}[\IEEEsetlabelwidth{12}]
	\item [1)]
	We propose a novel theory for accelerating SMPC task in the semi-honest adversary model. We find that the SMPC task of combining data from different parties using operations satisfying commutativity and associativity can be decomposed. Since local tasks do not use SMPC technology, they do not need to be calculated using Oblivious Transfer (OT) and Garbled Circuit (GC). So the computation speed of local tasks is much faster than SMPC tasks. For the same SMPC task, the decomposed SMPC task consumes less resources than the undecomposed SMPC task.
	\item [2)]
	We analyze the seurity of our theory. Besides, we list some examples that can be decomposed. In addition, we give the maximum number of times a SMPC task can be decomposed when there are $m$ parties participating in a SMPC task and each party has $n$ parameters, and we prove that under the maximum number, neither party can deduce the raw data from the decomposed SMPC task. We also provide methods to improve the security of decomposed SMPC tasks.
	\item [3)]
	We decompose three linear dimension reduction ML SMPC tasks, namely principal component analysis (PCA), singular value decomposition (SVD), and factor analysis (FA), to show the performance of our theory. These three tasks are more complex than the ML algorithms implemented by SMPC before, and they better show the performance of our theory. We use ABY \cite{Demmler2015ABYA} to implement these three tasks. Then we compare the resource consumption before and after the decomposition of these three tasks. The experimental results show that our theory is efficient. In addition, we use incremental computation technique to increase the amount of data involved in these three decomposed SMPC tasks. Finally, We compare the results of these three SMPC tasks with the results obtained by the Python standard library to prove that the results of these three SMPC tasks are accurate.

\end{enumerate}

\section{Preliminaries}

In this section, we introduce the notations and relevant knowledge.

\subsection{Notations}

We define several notations as follows:
\begin{enumerate}[\IEEEsetlabelwidth{12}]
	\item [1)] 
	$f_{m,n} = f(x_{11}, x_{12}, ..., x_{1n}, x_{21}, x_{22}, ..., x_{2n}, ..., x_{m1}\\, x_{m2}, ...,x_{mn}) = output$. The function $f_{m,n}$ means that there are $m$ parties to caculate this function together, and each party has $n$ parameters. $x_{ij}(1\le i \le m, 1\le j \le n)$ is the parameter which belongs to the $i$-th party. $output$ is the result of $f_{m,n}$, it can be a constant, a vector, a matrix, etc.
	\item [2)]
	$f_{n_i}^i$ means that this function belongs to $i$-th party and the function has $n_i$ parameters.
	\item [3)] 
	The $\odot$ is a operator that satisfies commutativity and associativity. For example, $\odot$ can represent the following operators: +, -, $\cdot$ (only for number), $\land$ , \& , $\lor$, $\oplus$, etc. 
	\item [4)] 
	$a, b, c, d, m, n, k$ are constants.
	\item [5)]
	$\mathbf{A_{m,n}}$ is a matrix with $m$ rows and $n$ columns.
	\item [6)]
	$\mathbf{A_{m,n}\top}$ is the transpose of $\mathbf{A_{m,n}}$.
\end{enumerate}

\subsection{Associativity and Commutativity}
In mathematics, the associative property is a property of some binary operations, which means that rearranging the parentheses in an expression will not change the result \cite{Associative}. For example:
\begin{enumerate}[\IEEEsetlabelwidth{12}]
	\item [1)]
	$(a + b) + c = a + (b + c)$;
	\item [2)]
	$(a \cdot b) \cdot c = a \cdot (b \cdot c)$.
\end{enumerate}
In mathematics, a binary operation is commutative if changing the order of the operands does not change the result\cite{Commutative}. For example:
\begin{enumerate}[\IEEEsetlabelwidth{12}]
	\item [1)]
	$a + b = b + a$;
	\item [2)]
	$a \cdot b = b \cdot a$.
\end{enumerate}

\subsection{Incremental Computation}
The main idea of incremental computation is to remove the calculated raw data from memory while retaining the calculation result, and then read a new batch of data for calculation until all data are calculated, and finally combine the results. The incremental computation technology can be used to complete the computing task when the memory capacity is not enough to accommodate all the raw data to complete a computing task. Following are the steps for incremental calculation:
\begin{enumerate}[\IEEEsetlabelwidth{12}]
	\item [1)]
	Divide the raw data into $n$ pieces.
	\item [2)]
	Read one piece of the raw data at a time for calculation and save the result.
	\item [3)] 
	Delete the raw data that have participated in calculations from memory.
	\item [4)] 
	Repeat reading the next piece data for calculation until all the raw data are calculated.
	\item [5)] 
	Merge all results.
\end{enumerate}
In Section \ref{section4}, we use incremental computation technique in Algorithm \ref{algorithm1}, Algorithm \ref{algorithm4} and Algorithm \ref{algorithm6} to increase the data involved in model training.

\subsection{Secure Multi-party Computation}
SMPC \cite{4568388} \cite{10.1145/3387108} \cite{Cramer2015SecureMC} \cite{Bayatbabolghani2013SecureMC} enables parties to perform distributed computing tasks in a secure manner. In other words, SMPC allows parties to use their private data to jointly participate a computation task without revealing their private data. SMPC is mainly implemented with OT \cite{Rabin2005HowTE}, GC \cite{4568207}, homomorphic encryption \cite{Gentry2009FullyHE}, secret sharing and other technologies. It has the two main security models: semi-honest adversary model and malicious attack model. SMPC is mainly used in machine learning security, privacy set computing, and secure genome sequence comparison, etc. SMPC should meet the following attributes:
\begin{enumerate}[\IEEEsetlabelwidth{12}]
	\item [1)]
	Privacy: Neither party can learn anything from other parties except the output. In particular, the only information that each party can learn is deduced from the output.
	\item [2)]
	Correctness: The output should be correct after the SMPC task is finished. This means each party can get correct output and no party can influence this.
	\item [3)]
	Independence of Inputs: Each party can choose their inputs independently.
	\item [4)] 
	Guaranteed Output Delivery: The corrupted parties cannot prevent the honest parties from obtaining the final output
	\item [5)] 
	Fairness: All parties should get the output, both corrupted parties and honest parties.
\end{enumerate}

\subsection{Oblivious Transfer}
OT is a protocol in cryptography. Based on OT, a sender can transmit one of many messages to a receiver without knowing which one was transmitted. While OT requires public-key cryptography, OT extensions \cite{Keller2015ActivelySO} \cite{10.1145/2508859.2516738} \cite{Kolesnikov2013ImprovedOE} use only a few basic OT and symmetric cryptography to speed up OT execution, such as Correlated OT (C-OT) \cite{10.1145/2508859.2516738} and Random OT (R-OT) \cite{10.1145/2508859.2516738}.

C-OT is suitable for secure computation protocols that require correlated inputs, such
as Free-XOR technique \cite{Kolesnikov2008ImprovedGC} for GC. 

R-OT is suitable for protocols where input are random numbers, such as using the GMW protocol \cite{10.1145/3335741.3335755} to generate beaver multiplication triples \cite{10.1007/3-540-46766-1_34}.

\subsection{Garbled Circuit}
GC \cite{4568207} is an encryption protocol that enables two parties to perform secure computations. In other words, two parties who do not trust each other can jointly compute functions on their private inputs without a trusted third party. The classic millionaire problem is that two millionaires want to compare who has more money, but neither wants to reveal the amount of their money. Yao proposed solving the millionaire problem through encrypted circuit in the 1980s, and since then there has been the concept of GC. If Alice want to build a GC, Alice should:
\begin{enumerate}[\IEEEsetlabelwidth{12}]
	\item [1)]
	Build the truth table for each gate.
	\item [2)]
	Choose random numbers to replace 0 and 1 of truth table and transform truth table into a "label table".
	\item [3)]
	Encrypt the output label with the input label as keys to a double encryption.
\end{enumerate}
If Alice want to compute the GC with Bob, Alice should:
\begin{enumerate}[\IEEEsetlabelwidth{12}]
	\item [1)]
	Send the GC, scrambled outputs and his inputs to Bob.
	\item [2)]
	Send Bob's inputs to Bob using OT.
	\item [3)]
	Wait for Bob to compute the GC.
	\item [4)]
	Get the output from Bob and share the result.
\end{enumerate}
This website \cite{Garbled-circuit-protocol} has a more detailed introduction.
\subsection{Security Model}
Our theory is based on the \textbf{semi-honest adversary model}.
In the semi-honest adversary model, there are honest parties and corrupted parties, and they will abide by the execution of the agreement. However, the corrupted parties try to infer the raw data of other parties from the information they have already obtained. Although the semi-honest adversary model is a weak adversarial model, there are still many application scenarios in the real world.

\subsection{Machine Learning}
In this section, we briefly describe the three ML linear dimension reduction algorithms covered in this paper: PCA \cite{Principal-component-analysis}, SVD \cite{Golub2007SingularVD} and FA \cite{Harman1960ModernFA}.
\subsubsection{Principal Component Analysis}
	PCA is the most widely used dimension reduction algorithm. The main idea of PCA is to map $n$-dimensional features to $k$-dimensional features, which are new orthogonal features, also known as principal components.

	Suppose we have a dataset containing $m$ pieces of data, each with $n$ attributes, we can form this dataset into a matrix $\mathbf{A_{m,n}}$. Then we can use the PCA to first find the eigenvalues and eigenvector matrices $\mathbf{V_{n,n}}$ corresponding to the covariance matrix of $\mathbf{A_{m,n}}$. Then select the eigenvectors corresponding to the largest $k$ eigenvalues and form $\mathbf{V_{n,k}}$, and finally compute $\mathbf{A_{m,k}} = (\mathbf{A_{m,n}} - \mathbf{AVG_{m, n}}) \cdot \mathbf{V_{n,k}}$ to reduce the number of attributes of the data to $k$.

	PCA is used to reduce data noise while retaining as much original data information as possible. Besides, PCA is also used to preserve the data privacy by converting $n$-demensional data into $k$-demansional data.
\subsubsection{Singular Value Decomposition}
	SVD decomposes the matrix of any $m$ rows and $n$ columns into the multiplication of three sub matrices.

	If we have a dataset containing $m$ pieces of data, each with $n$ attributes, we can form this dataset into a matrix $\mathbf{A_{m,n}}$. Then we can use the SVD method to decompose $\mathbf{A_{m,n}}$ into $\mathbf{U_{m,m}} \cdot \mathbf{\sum_{m,n}} \cdot \mathbf{V_{n,n}^\top}$. If we want to compress the rows of $\mathbf{A_{m,n}}$, we can take the first $k$ rows of $\mathbf{U_{m,m}^\top}$ to calculate $\mathbf{A_{k,n}} = \mathbf{U_{k,m}^\top} \cdot \mathbf{A_{m,n}}$, and if we want to compress the columns of $\mathbf{A_{m,n}}$ , we can take the first $k$ columns of $\mathbf{V_{n,k}}$ to calculate $\mathbf{A_{m,k}} = \mathbf{A_{m,n}} \cdot \mathbf{V_{n,k}}$.

	SVD is not only used in image compression, digital watermarking, recommendation system and article classification, but also has good applications in signal decomposition, signal reconstruction, signal noise reduction, etc. SVD is the cornerstone of many ML algorithms.
\subsubsection{Factor Analysis}
	FA is a statistical technique to extract common factors from variable groups. When we want to understand a phenomenon, we often evaluate and measure it through multi-dimensional indicators. However, data collection and analysis will become difficult when there are too many dimension indicators. To make data collection and analysis easy, we need to use FA to reduce the number of attributes of the data.

	If we have a dataset containing $m$ pieces of data, each with $n$ attributes, then we can form this dataset into a matrix $\mathbf{A_{m,n}}$. We first calculate the correlation coefficient matrix $\mathbf{B_{n,n}}$ of $\mathbf{A_{m,n}}$, and then calculate the eigenvalues and eigenvector matrix of $\mathbf{B_{n,n}}$. Then select the largest $k$ eigenvalues as the principal factors, and then convert the corresponding $k$ eigenvectors into a factor load matrix. The factor load matrix is then rotated namely $\mathbf{F_{n, k}}$. Finally, $\mathbf{F_{n, k}}$ is used to reduce the dimensionality of the original data.

\section{SMPC Task Decomposition}

In this section, we introduce our theory, called SMPC Task Decomposition (SMPCTD). SMPCTD can be used to decompose these SMPC tasks that use $\odot$ to combine data from diffrent parties. The resources consumed by the decomposed SMPC tasks are not only reduced but also stabilized within a certain range. First, we briefly introduce our theory in Section \ref{section3.1}. Second, we present our theory in detail and list some examples that can be decomposed in Section \ref{section3.2}. Third, we analyze the security of our theory in Section \ref{section3.3}. Finally, we introduce methods to enhance the security of decomposed SMPC tasks in Section \ref{section3.4}. Note that our theory is only used in the semi-honest adversary model, and the decomposed SMPC task is executed only once.

\subsection{Theory Overview}
\label{section3.1}
We find that in ML algorithms, there are a large number of operations satisfying commutability and associativity. These operations often require the participation of large amount of raw data. If we generate a GC for each raw data, it will consume a lot of memory and bring a great burden to communication. In addition, heavy use of OT consumes a lot of time.

Our theory stems from the fact that $\odot$ can be decomposed into combinations of multiple operations. For example, addition satisfies commutability and associativity, if $f(x, y) = ax + by$, where $a$ and $b$ are constants, we can calculate $ax$ and $by$ respectively and then add them together.

Based on this property, we can decompose a single SMPC task which uses $\odot$ to combine data from diffrent parties into multiple SMPC sub-tasks and multiple local tasks. Thus, each party can compute partial result locally and combine all results using SMPC sub-task. Computing in this way can greatly reduce the consumption of time, memory and communication without revealing the raw data. 

\subsection{Theory Details}
\label{section3.2}
In this section, we introduce the details of our theory:
\begin{enumerate}[\IEEEsetlabelwidth{12}]
	\item [1)]
	\textbf{A SMPC task which uses $\odot$ to combine data from diffrent parties can be decomposed into one SMPC sub-task and one local task}. We describe this theory with an equation: $f_{m,n} = f_{n}^1 \odot f_{n}^2  \odot ...\odot f_{n}^i\odot... \odot f_{n}^m$. $i$-th party can locally compute $f_{n}^i = output_i$. Then parties can combine $ouput_1$ to $output_m$ through one SMPC sub-task.
	
	We list some examples that can be decomposed:
	\begin{enumerate}[\IEEEsetlabelwidth{12}]
		\item [a)] $f_{m,n} = \sum_{i=1}^{m}\sum_{j=1}^nc(ax_{ij} + b)^k$
		\item [b)] $f_{m,n} = \sum_{i=1}^{m}\sum_{j=1}^n\log(c(ax_{ij} + b)^k)$
		\item [c)] $f_{m,n} = \exp(\sum_{i=1}^{m}\sum_{j=1}^nc(ax_{ij} + b)^k)$
		\item [d)] $f_{m,n} = \mathbf{A_{mn,d}}^\top\cdot\mathbf{A_{mn,d}}$
	\end{enumerate}
	We explain d). Considering that there are $m$ parties, each party has $n$ pieces of data. Each data has $d$ attributes and is represented as a row vector, e.g. $x = (y_1, y_2, ..., y_d)$. Parties can combine all data into $\mathbf{A_{mn,d}}$. Suppose parties want to compute $\mathbf{A_{mn,d}^\top} \cdot \mathbf{A_{mn,d}}$. Parties should put all data into a SMPC task and then calculate $\mathbf{A_{mn,d}^\top} \cdot \mathbf{A_{mn,d}}$ according to the traditional SMPC calculation method. However, this process requires a lot of time, memory and communication. If parties break this SMPC task down, they don't have to combine the data into $\mathbf{A_{mn,d}}$ for calculation. Instead, they can calculate the $\mathbf{A_{n,d}^\top} \cdot \mathbf{A_{n,d}}$ separately and add the results together. For example:
	\begin{eqnarray}
		\mathbf{A_{4,4}} &=& \left(\begin{IEEEeqnarraybox*}[][c]{,c/c/c/c,}
			1&2&3&4\\
			5&6&7&8\\
			9&10&11&12\\
			13&14&15&16%
		\end{IEEEeqnarraybox*}\right)\nonumber\\
		\mathbf{A_{4,4}^\top} \cdot \mathbf{A_{4,4}}
		&=& \left(\begin{IEEEeqnarraybox*}[][c]{,c/c/c/c,}
			276&304&332&360\\
			304&336&368&400\\
			332&368&404&440\\
			360&400&440&480%
		\end{IEEEeqnarraybox*}\right)\nonumber
	\end{eqnarray}
	
	$\mathbf{A_{4,4}}$ can be divided into $\mathbf{A_{2, 4}^1}$ and $\mathbf{A_{2, 4}^2}$:
	\begin{eqnarray}
		\mathbf{A_{2, 4}^1} &=& \left(\begin{IEEEeqnarraybox*}[][c]{,c/c/c/c,}
			1&2&3&4\\
			5&6&7&8%
		\end{IEEEeqnarraybox*}\right)\nonumber\\
		\mathbf{A_{2, 4}^2} &=& \left(\begin{IEEEeqnarraybox*}[][c]{,c/c/c/c,}
			9&10&11&12\\
			13&14&15&16%
		\end{IEEEeqnarraybox*}\right)\nonumber
	\end{eqnarray}
	
	 Then $\mathbf{A_{2, 4}^{1\top}} \cdot \mathbf{A_{2, 4}^1}$ and $ \mathbf{A_{2, 4}^{2\top}} \cdot \mathbf{A_{2, 4}^2}$ are calculated separately:
	\begin{eqnarray}
		\mathbf{A_{2,4}^{1\top}} \cdot \mathbf{A_{2,4}^1} &=& \left(\begin{IEEEeqnarraybox*}[][c]{,c/c/c/c,}
			26& 32& 38& 44\\
			32& 40& 48& 56\\
			38& 48& 58& 68\\
			44& 56& 68& 80%
		\end{IEEEeqnarraybox*}\right)\nonumber\\
		\mathbf{A_{2,4}^{2\top}} \cdot \mathbf{A_{2,4}^2} &=& \left(\begin{IEEEeqnarraybox*}[][c]{,c/c/c/c,}
			250&272&294&316\\
			272&296&320&344\\
			294&320&346&372\\
			316&344&372&400%
		\end{IEEEeqnarraybox*}\right)\nonumber
	\end{eqnarray}

	Finally $\mathbf{A_{4,4}^\top} \cdot \mathbf{A_{4,4}} = \mathbf{A_{2,4}^{1\top}} \cdot \mathbf{A_{2,4}^1} + \mathbf{A_{2,4}^{2\top}} \cdot \mathbf{A_{2,4}^2}$
	\item [2)]
	\textbf{The number of times a SMPC task can be decomposed is equal to the number of times $\odot$ is used to combine data from all parties}. For example, if $f_{m,n} = \sum_{i=1}^{m}\sum_{j=1}^{n}(x_{ij}-\bar{x})$, $f_{m,n}$ can be decomposed twice:
	\begin{equation*}
		\begin{cases}
			\bar{x} = \frac{1}{mn} \cdot \sum_{j=1}^{n}x_{1j} \odot ... \odot \frac{1}{mn}  \sum_{j=1}^{n}x_{mj}\\
			f_{m,n} = \sum_{j=1}^{n}(x_{1j}-\bar{x}) \odot ... \odot \sum_{j=1}^{n}(x_{mj}-\bar{x})
		\end{cases}
	\end{equation*}
	\item [3)]
	\textbf{The raw data must cannot be inferred from the output of SMPC sub-task}. After the decomposition, the results calculated by all parties through local task will be combined through a SMPC sub-task. The output of the SMPC sub-task needs to be returned to all parties. Therefore, the output of SMPC sub-task must not reveal the raw data of any parties.
	
	For example, if two parties want to calculate the average of their money, this is not allowed because either party can deduce the amount of money the other party has by the equation $ x + y = 2 \cdot average\_money$. But if three parties want to calculate the average of their money, this is allowed.
	\item [4)]
	\textbf{If one SMPC sub-task generates one equation, there are at most $[n \cdot (m-1) - 1]$ numbers of SMPC sub-tasks in cases where there are $m$ parties and each party has $n$ parameters}. If a SMPC task is decomposed into $n \cdot (m-1)$ SMPC sub-tasks, each party will get $n \cdot (m-1)$ equations. If the $n \cdot (m-1)$ equations are linear uncorrelated, then either party can deduce $n \cdot (m-1)$ unknowns from the $n \cdot (m-1)$ equations. For example, considering the equation system \ref{equ1}:
	\begin{equation}
		\begin{cases}
			\label{equ1}
			a_{11}x_{1} + a_{12}x_{2} + a_{13}x_{3} + a_{14}x_{4} = c_{1} \\
			c_{1} \cdot (a_{21}x_{1} + a_{22}x_{2} + a_{23}x_{3} + a_{24}x_{4}) = c_{2}
		\end{cases}
	\end{equation}
	 There are four parties and each party has one parameter in equation system \ref{equ1}. There can only be a maximum of $[1 \cdot (4 - 1) - 1] = 2$ equations according to our theory, so equation system \ref{equ1} does not reveal unknowns on either party. But if there is one more linear uncorrelated equation, either party can infer the unknowns of the others.
	\item [5)]
	\textbf{The final output must be computed in the last SMPC sub-task}. This rule ensures that each party receives the same result.

\end{enumerate}

\subsection{Security Analysis}
\label{section3.3}
In this section, we first analyze the security of our theory in Section \ref{section3.3.1}. Then we compare the security of the decomposed SMPC task with the security of the traditional SMPC task in Section \ref{section3.3.2}.
\subsubsection{Theory Security Analysis}
\label{section3.3.1}
We start by analyzing the security of each task. We decompose a SMPC task into multiple SMPC sub-tasks and multiple local tasks according to our theory. For each local task, each party computes it locally without any communication. Therefore, neither party can access the private data of others. For each SMPC sub-task, all parties calculate the output together through a SMPC sub-task, and neither party can deduce the raw data from the output according to the third rule of our theory.

We then combine all the tasks into a whole task to analyze security. Local tasks are always secure because all parties compute local tasks in their own secure environment without any communication. Let us analyze the security of the entire decomposed SMPC task. If there are $m$ parties and each party has $n$ parameters, a SMPC task is decomposed up to $[n \cdot (m-1) - 1]$ SMPC sub-tasks. Thus, either party can only get $[n \cdot (m-1) - 1]$ equations at most, while the $[n \cdot (m-1) - 1]$ equations cannot deduce $n \cdot (m-1)$ unknowns. So the whole task is secure.

\subsubsection{Security Compared to Traditional SMPC Task}
\label{section3.3.2}
Traditional SMPC tasks require waiting until all data preparation is complete before starting calculation. Therefore, each party can only get one equation, which is $f_{m,n} = output$. Suppose a SMPC task is decomposed into $k$ SMPC sub-tasks and several local tasks according to our theory. Each party can get the following $k$ equations:
\begin{equation*}
	k
	\begin{cases}
		f_{m,n}^1 = output_1 \\
		f_{m,n}^2 = output_2 \\ 
		... \\
		f_{m,n}^k = output_k
	\end{cases}
\end{equation*}

However, as long as $k$ is less than $[n \cdot (m-1)]$, the decomposed SMPC task is safe.
For example, if $m$ parties want to calculate $f_{m,n} = \sum_{i=1}^{m}\sum_{j=1}^n(x_{ij} - \bar{x})^2$, they can calculate the result according to the following steps:
\begin{enumerate}[\IEEEsetlabelwidth{12}]
	\item[1)] 
	The $i$-th party locally computes $s_i = \sum_{j=1}^n x_{ij}$. 
	\item[2)] 
	Then all parties compute $ \bar{x} = \frac{1}{mn}\sum_{i=1}^{m}s_i$ by a SMPC sub-task. 
	\item[3)] 
	The $i$-th party locally computes  $sum_i = \sum_{j=1}^n (x_{ij} - \bar{x})^2$.
	\item[4)] 
	Then all parties compute $ output = \sum_{i=1}^m sum_i$ by a SMPC sub-task.
\end{enumerate}
The above example is to decompose one SMPC task into two SMPC sub-tasks and two local tasks. So each party can get the following two equations:
\begin{equation*}
	\begin{cases}
		\frac{1}{mn}\sum_{i=1}^{m}\sum_{j=1}^{n}x_{ij} =\bar{x} \\
		\sum_{i=1}^{m}\sum_{j=1}^n(x_{ij} - \bar{x})^2 = output
	\end{cases}
\end{equation*}
And neither party can derive the raw data of the other party as long as $n \cdot (m-1) > 2$. 
\subsection{Methods to Enhance the Security of Decomposed SMPC Tasks}
\label{section3.4}
\subsubsection{Reduce the Number of Decompositions}
This method is the easiest way to enhance security of decomposed SMPC tasks. The fewer times a SMPC task is decomposed, the less information each party gets, and the safer the SMPC task is. When the number of decompositions is zero, the decomposed task becomes a traditional task.

\subsubsection{Add Unknowns to Each SMPC Sub-task}
\label{section3.4.2}
For example, if $m$ parties want to calculate the mean of their data, and each party has different amounts of data, such as $n_1, n_2, ..., n_m$. Then $i$-th party can put $n_i$ as unknown into the SMPC sub-task. Finally, each party can get the following equation:
\begin{equation*}
	\sum_{i=1}^{m}\frac{n_i}{(n_1 + n_2 +...+n_m)}\cdot\frac{1}{n_i}\sum_{j=1}^{n_i}x_{ij} =\bar{x}
\end{equation*}
As a result, neither party has access to the sum and amount of data from any other party. We use this method in Algorithm \ref{algorithm2}, Algorithm \ref{algorithm3} and Algorithm \ref{algorithm7}.

\subsubsection{Make Irreversible Changes to the Output}
\label{section3.4.3}
In other words, if the output of the SMPC sub-task is irreversibly transformed, the number of equations obtained by each party is reduced by one. For example, if a vector is normalized, then we cannot restore it to the original vector. We use this method in Algorithm \ref{algorithm4} and Algorithm \ref{algorithm7}.

\section{Three Decomposed ML SMPC Tasks}
\label{section4}

In this section, we use our theory to decompose three linear dimension reduction ML SMPC tasks. The three ML SMPC tasks are PCA, SVD and FA. We present the specific algorithms and conduct security analysis of the algorithms. We present the decomposed PCA SMPC task in Section \ref{section4.1}. Then we present the decomposed SVD SMPC task in Section \ref{section4.2}. Finally, we present the decomposed FA SMPC task in Section \ref{section4.3}

\subsection{Decomposed PCA SMPC task}
\label{section4.1}
\subsubsection{Algorithm}
Algorithm \ref{algorithm1} describes the decomposed PCA SMPC task. The inputs of Algorithm \ref{algorithm1} are the path name of the datasets and the number of datasets. The outputs of Algorithm \ref{algorithm1} are the mean of the data, the eigenvalues and eigenvectors corresponding to the covariance matrix of data.

In Algorithm \ref{algorithm1}, lines 2 through 6 read the data incrementally and save the sum of the data in $column\_sum$. Line 7 calculates the average of the $column\_sum$. Line 8 calculates the average of the total data from all parties through a SMPC sub-task. Lines 9 through 14 calculate the covariance matrix incrementally. In line 15, the SMPC sub-task first merges the covariance matrices of all parties, and then calculates the eigenvalues and eigenvectors of the covariance matrix. Once the trained model is received, parties can freely reduce the dimensions of their data.

Algorithm \ref{algorithm2} is a SMPC sub-task that calculates the mean of all data. The inputs of Algorithm \ref{algorithm2} are the average of each party's data and the amount of each party's data. The output is the average of all data. We use the method described in Section \ref{section3.4.2} to hide the quantity and mean of data from all parties.

Algorithm \ref{algorithm3} is the power iteration method \cite{Golub1983MatrixC} for calculating the eigenvalues and eigenvectors of a matrix. In line 13, we use the Rayleigh quotient to speed up the convergence of eigenvalues. If we use Algorithm \ref{algorithm3} as a SMPC task, this algorithm first merges the matrices of all parties together.
\subsubsection{Security Analysis}
We decompose the PCA SMPC task into two SMPC sub-tasks, which are Algorithm \ref{algorithm2} and Algorithm \ref{algorithm3}. Algorithm \ref{algorithm2} does not expose the mean and quantity of data of each party. Algorithm \ref{algorithm3} also does not expose the inputs. So Algorithm \ref{algorithm2} and Algorithm \ref{algorithm3} are secure. Finally, each party can obtain the following equation system through Algorithm \ref{algorithm1}:
\begin{equation*}
	\begin{cases}
		\sum_{i=1}^{m}\frac{n_i}{(n_1 + n_2 +...+n_m)}\cdot\frac{1}{n_i}\sum_{j=1}^{n_i}x_{ij} =\bar{x} \\
		
		\sum_{i=1}^{m} \frac{n_i}{(n_1 + n_2 +...+n_m - 1)}matrix_i = matrix
	\end{cases}
\end{equation*}
When the number of unknowns is greater than two, neither party can deduce the unknowns of the other party. Obviously the number of data in the dataset is greater than two, so algorithm \ref{algorithm1} is secure.
\begin{algorithm}
	\label{algorithm1}
	\SetAlgoLined
	\caption{Decomposed PCA SMPC task}
	\KwIn{$file\_name$, $dataset\_number$}
	\KwOut{$total\_avg, eigenvalue\_array, \newline
		eigenvector\_matrix$}
	\textbf{extern} ReadData, CptTotColAvg, PowerIteration;\\
	\For{i :=1 to dataset\_number}
	{
		$temp\_data$ $\leftarrow$ ReadData($file\_name\_i$);\\
		$total\_lines$ $\leftarrow$ $total\_lines + \newline temp\_data.size()$;\\
		$column\_sum$ $\leftarrow$ $column\_sum + temp\_data$;
	}
	$col\_avg$ $\leftarrow$ $column\_sum / total\_lines$;\\
	$total\_avg$ $\leftarrow$ CptTotColAvg($col\_avg, total\_lines$);// SMPC task\\
	\For{i :=1 to dataset\_number}
	{
		$temp\_data$ $\leftarrow$ ReadData($file\_name\_i$);\\
		$temp\_data$ $\leftarrow$ $temp\_data - total\_avg$;\\
		$temp\_cov\_mat$ $\leftarrow$ $temp\_data\_t \cdot temp\_data$;\\
		$cov\_mat$ $\leftarrow$ $cov\_mat + \newline temp\_cov\_mat / total\_lines$;
	}
	$eigenvalue\_array, eigenvector\_matrix$ $\leftarrow$ PowerIteration($cov\_mat$); // SMPC task
\end{algorithm}\\
\begin{algorithm}
	\label{algorithm2}
	\SetAlgoLined
	\caption{CptTotColAvg}
	\KwIn{$col\_avg_i$, $n_i$}
	\KwOut{$total\_avg$}
	$total\_avg$ $\leftarrow$ $\sum_{i=1}^{m}$$\frac{n_i}{(n_1 + n_2 +...+n_m)}col\_avg_i$;
\end{algorithm}\\
\begin{algorithm}
	\label{algorithm3}
	\SetAlgoLined
	\caption{PowerIteration}
	\KwIn{$matrix_m$, $n_i$}
	\KwOut{$eigenvalue\_array, eigenvector\_matrix$}
	\eIf{use this algorithm as a SMPC task}
	{
		$matrix$ $\leftarrow$ $\sum_{i=1}^{m}$$\frac{n_i}{n_1 + n_2 +...+n_m - 1}matrix_i$;
	}
	{
		$matrix$ $\leftarrow$ $matrix_m$;
	}
	
	\For{$i := 0$ to $(matrix.columns() - 1)$}
	{
		$v \leftarrow matrix[\;][0]$;\\
		\For{$j:=0$ to $iteration\_round$}
		{
			$y$ $\leftarrow$ $v / \Vert v\Vert$;\\
			$v$ $\leftarrow$ $matrix$ $\cdot$ $y$;
		}
		$y\_t$ $\leftarrow$ $y.transpose()$;\\
		$eigenvalue$ $\leftarrow$ $y\_t$ $\cdot$ $v$;\\
		$matrix$ $\leftarrow$ $matrix - eigenvalue$ $\cdot$ $y$ $\cdot$ $y\_t$;\\
		$eigenvalue\_array[i]$ $\leftarrow$ $eigenvalue$;\\
		$eigenvector\_matrix[\;][i]$ $\leftarrow$ $y$;
	}
\end{algorithm}
\subsection{Decomposed SVD SMPC task}
\label{section4.2}
\subsubsection{Algorithm}
Algorithm \ref{algorithm4} describes the decomposed SVD SMPC task. The purpose of the SVD task is to obtain singular values and the right singular matrix. Because we arrange the attributes of the data in columns, we only need the right singular matrix to reduce the dimensions of the columns. If we arrange the attributes of the data in rows, we need to return the left singular matrix.

In Algorithm \ref{algorithm4}, the inputs are the path name and the number of datasets. The outputs are the singular values and the right singular matrix of all the data from parties. Lines 2 through 6 read data incrementally and incrementally calculate the transpose of the total data matrix multiplied by the total data matrix. In line 7, the SMPC sub-task first combines the data of all parties, then calculates the singular values and the right singular matrix of the total data. Finally, this algorithm returns singular values and a right singular matrix so that all parties can freely reduce the dimensions of their data.

In Algorithm \ref{algorithm5}, The input is the $i$-party's matrix. The outputs are the singular values and the right singular matrix of the total data. Line 3 merges the matrices of all parties together. Line 4 uses power iteration method to calculate the eigenvalues and the right singular matrix. Line 5 computes the singular values.
\subsubsection{Security Analysis}
We decompose the SVD SMPC task into one SMPC sub-task and one local task. Theoretically, each party can get one equation:
\begin{equation}
	\sum_{i=1}^{m}matrix_i = matrix \nonumber
\end{equation}
But we do not return the left singular matrix, so no party can recover the $matrix$ through $\mathbf{U} \cdot \mathbf{\sum} \cdot \mathbf{V^\top}$. Thus neither party can get even one equation. The security of Algorithm \ref{algorithm4} is the same as that of the traditional SVD SMPC task, but the resource consumption of Algorithm \ref{algorithm4} is much lower than that of the traditional SVD SMPC task.
\begin{algorithm}
	\label{algorithm4}
	\SetAlgoLined
	\caption{Decomposed SVD SMPC task}
	\KwIn{$file\_name, dataset\_number$}
	\KwOut{$singular\_value, right\_singular\_matrix$}
	\textbf{extern} ReadData, ComputeSingularValueAndMatrix;\\
	\For{$i := 1$ to $dataset\_number$}
	{
		$temp\_data$ $\leftarrow$ ReadData($file\_name\_i$);\\
		$temp\_data\_t$ $\leftarrow$ $temp\_data.transpose()$;\\
		$data\_matrix$ $\leftarrow$ $data\_matrix + temp\_data\_t$ $\cdot$ $temp\_data$;
	}
	
	$singular\_value, right\_singular\_matrix$ $\leftarrow$ ComputeSingularValueAndMatrix($data\_matrix$); // SMPC task
\end{algorithm}
\begin{algorithm}
	\label{algorithm5}
	\SetAlgoLined
	\caption{ComputeSingularValueAndMatrix}
	\KwIn{$matrix_i$}
	\KwOut{$singular\_value, right\_singular\_matrix$}
	\textbf{extern} PowerIteration;\\
	// This is a SMPC Task , so we first merge the matrices of all parties together.\\
	$matrix$ $\leftarrow$ $\sum_{i=1}^{m} matrix_i$\\
	$eigenvalue, right\_singular\_matrix$ $\leftarrow$ PowerIteration($matrix$);\\
	$singular\_value$ $\leftarrow$ $\sqrt{eigenvalue}$;
\end{algorithm}
\subsection{Decomposed FA SMPC task}
\label{section4.3}
\subsubsection{Algorithm}
Algorithm \ref{algorithm6} describes the decomposed FA SMPC task. The purpose of the FA task is to obtain the principal factors and factor load matrix of the data. With the principal factors and factor loading matrices, each party can freely choose an appropriate number of principal factors as needed, and further rotate the factor loading matrix for analyzing their data.

In Algorithm \ref{algorithm6}, the inputs are the path name and the number of datasets. The outputs are the principal factors and factor load matrix of all the data. Lines 2 through 6 read the data incrementally, and save the sum of the data in $column\_sum$. Line 7 calculates the average of the total data. Line 8 averages the data for all the parties through a SMPC sub-task. Lines 9 through 14 incrementally calculate the covariance matrices of each party. In line 15, the SMPC sub-task first combines the covariance matrices of each party , then calculates the correlation coefficient matrix, and finally calculates the principal factors and the factor loading matrix.

In Algorithm \ref{algorithm7}, the two inputs are the $i$-th party's covariance matrix and the amount of data. The outputs are the principal factors and the factor load matrix of all data. Line 3 merges the covariance matrices of all parties together. lines 4 to 6 calculate the standard deviation of the raw data. Lines 7 through 11 calculate the correlation coefficient matrix of the raw data. line 12 uses power iteration mathod to calculate the exact eigenvalues of the correlation coefficient matrix. Line 13 uses the shift power iteration method to accelerate the calculation of the eigenvectors of the correlation coefficient matrix. Lines 14 through 18 calculate the factor loading matrix of the raw data.

In Algorithm \ref{algorithm8}, the inputs are matrix and the eigenvalues of the matrix. Because the shift power iteration method needs the exact eigenvalues of the matrix. The outputs are the eigenvalues and the eigenvectors of the input matrix. One disadvantage of the power iteration method is that the convergence of eigenvectors is very slow when the maximum eigenvalue and the second eigenvalue of the matrix are close. So we use the shift method to accelerate the convergence speed of the eigenvectors.
\subsubsection{Security Analysis}
In Algorithm \ref{algorithm6}, we decompose the FA SMPC task into two SMPC sub-tasks. However, we use the method described in Section \ref{section3.4.3} to enhance the security of Algorithm \ref{algorithm7}, we merge the covariance matrices of the parties into $matrix$ and convert $matrix$ to a correlation coefficient matrix called $correlation\_matrix$, and this conversion is irreversible. So each party can only get one equation:
\begin{equation}
	\label{eqution3}
	\sum_{i=1}^{m}\frac{n_i}{(n_1 + n_2 +...+n_m)}\cdot\frac{1}{n_i}\sum_{j=1}^{n_i}x_{ij} =\bar{x}
\end{equation}	
Thus neither party can infer the other's data when the unknown quantity of equation \ref{eqution3} is greater than one. Obviously, the number of unknowns must be greater than one as long as there are two parties involved in the calculation.

\begin{algorithm}
	\label{algorithm6}
	\SetAlgoLined
	\caption{Decomposed FA SMPC task}
	\KwIn{$file\_name, dataset\_number$}
	\KwOut{$principal\_factors, factor\_loading\_matrix$}
	\textbf{extern} ReadData, CptTotColAvg, ComputeFactorLoadingMatrix;\\
	\For{$i :=1$ to $dataset\_number$}
	{
		$temp\_data$ $\leftarrow$ ReadData($file\_name\_i$);\\
		$total\_lines$ $\leftarrow$ $total\_lines + \newline temp\_data.size()$;\\
		$column\_sum$ $\leftarrow$ $column\_sum + temp\_data$;
	}
	$col\_avg$ $\leftarrow$ $column\_sum / total\_lines$;\\
	$total\_avg$ $\leftarrow$ CptTotColAvg($col\_avg$, $total\_lines$);// SMPC task\\
	\For{$i := 1$ to $dataset\_number$}
	{
		$temp\_data$ $\leftarrow$ ReadData($file\_name\_i$);\\
		$temp\_data$ $\leftarrow$ $temp\_data - total\_avg$;\\
		$temp\_cov\_mat$ $\leftarrow$ $temp\_data.transpose() \cdot \newline temp\_data$;\\
		$cov\_mat$ $\leftarrow$ $cov\_mat + \newline temp\_cov\_mat / total\_lines$;
	}
	$principal\_factors, factor\_loading\_matrix$ $\leftarrow$ ComputeFactorLoadingMatrix($cov\_mat$, $total\_lines$); // SMPC task
\end{algorithm}
\begin{algorithm}
	\label{algorithm7}
	\SetAlgoLined
	\caption{ComputeFactorLoadingMatrix}
	\KwIn{$matrix_i$, $n_i$}
	\KwOut{$principal\_factors, factor\_loading\_matrix$}
	\textbf{extern} PowerIteration, ShiftPowerIteration;\\
	// This is a SMPC Task , so we first merge the matrices of all parties together.\\
	$matrix$ $\leftarrow$ $\sum_{i=1}^{m}$$\frac{n_i}{n_1 + n_2 +...+n_m-1}matrix_i$;\\
	\For{$i := 0$ to $(matrix.column\_size()-1)$}
	{
		$s\_d[i]$ $\leftarrow$ $\sqrt{matrix[i][i]}$
	}
	
	\For{$i := 0$ to $(matrix.row\_size()-1)$}
	{
		\For{$j := 0$ to $(matrix.column\_size()-1)$}
		{
			$correlation\_matrix[i][j]$ $\leftarrow$ $matrix[i][j] / (s\_d[i] \cdot s\_d[j])$;\\
		}
	}
	$eigenvalue\_array, temp\_matrix$ $\leftarrow$ PowerIteration($correlation\_matrix$);\\
	
	$principal\_factors, eig\_m$ $\leftarrow$ ShiftPowerIteration($correlation\_matrix,\newline eigenvalue\_array$);\\
	
	\For{$i := 0$ to $(eig\_m.row\_size()-1)$}
	{
		\For{$j := 0$ to $(eig\_m.column\_size()-1)$}
		{
			$factor\_loading\_matrix[i][j]$ $\leftarrow$ $eig\_m[i][j]$ $\cdot$ $\sqrt{principal\_factors[i]}$;
		}
	}
\end{algorithm}
\begin{algorithm}
	\label{algorithm8}
	\SetAlgoLined
	\caption{ShiftPowerIteration}
	\KwIn{$matrix$, $last\_eigenvalue\_array$}
	\KwOut{$eigenvalue\_array, eigenvector\_matrix$}
	
	\For{$i := 0$ to $(matrix.columns() - 1)$}
	{
		$v \leftarrow matrix[\;][0]$;\\
		\eIf{$i == (matrix.columns() - 1)$}
		{
			$alp \leftarrow last\_eigenvalue\_array[i] / 2$;
		}
		{
			$alp \leftarrow last\_eigenvalue\_array[i+1] / 2$;
		}
		$matrix\_B$ $\leftarrow$ $matrix - alp \cdot I$;\\
		\For{$j:=0$ to $iteration\_round$}
		{
			$y$ $\leftarrow$ $v / \Vert v\Vert$;\\
			$v$ $\leftarrow$ $matrix\_B$ $\cdot$ $y$;
		}
		$y\_t$ $\leftarrow$ $y.transpose()$;\\
		$eigenvalue$ $\leftarrow$ $y\_t$ $\cdot$ $v$ + $alp$;\\
		$matrix$ $\leftarrow$ $matrix - eigenvalue$ $\cdot$ $y$ $\cdot$ $y\_t$;\\
		$eigenvalue\_array[i]$ $\leftarrow$ $eigenvalue$;\\
		$eigenvector\_matrix[\;][i]$ $\leftarrow$ $y$;
	}
\end{algorithm}
\section{Evaluation \& Discussion}
\label{section5}
We implement the three decomposed ML SMPC tasks described in Section \ref{section4}. In this section, we present the experimental results. In Section \ref{section5.1}, we compare the resources consumed by the three tasks before and after the decomposition. In Section \ref{section5.2}, we further increase the data involved in the three decomposed tasks, and observe the consumption of time, memory and traffic. In Section \ref{section5.3}, we analyse the accuracy of the results of the three decomposed ML SMPC tasks. In Section \ref{section5.4}, we analyse the experimental results. 

\textbf{The Implementation}. We implement the three ML SMPC tasks based on ABY and the language is C++. In our experiments, the field type is set to 32-bit single-precision floating-point. Therefore, we mainly use the Boolean circuit in ABY. Besides, we use floating-point gates that are generated and optimized with hardware analysis tools\cite{10.1145/2810103.2813678}.

\textbf{Experimental settings}. We run two processes on our local machine. Our experiments are performed on a localhost with a CPU of intel corei7 11700k 3.6GHZ, 128GB of RAM, an Ubuntu 22.04.1 LTS operating system, a SSD for a hard drive, and a GeForce RTX 3080Ti graphics card. $n$ is the  data volume on one party, so the sum of the data volumes of both parties is $2n$.  

\textbf{Data sets}. In our experiments, the following two datasets are mainly used. The Wilt Data Set \cite{TheWiltDataSet} contains data from a remote sensing study by Johnson et al. (2013) that involved detecting diseased trees in Quickbird imagery. It has 4339 training samples, each with 6 attributes representing some of the basic characteristics of the tree. However, in order to facilitate our experiments, we have removed the categories of the data set. The Accelerometer Data Set \cite{TheAccelerometerDataSet} from vibrations of a cooler fan with weights on its blades. It has 153000 samples, each with 5 attributes representing the basic information of the cooler fan.

\subsection{Experiments Compared to Three Traditional SMPC Tasks}
\label{section5.1}
In this section, we compare the resources consumed by the three traditional SMPC tasks and the three decomposed SMPC tasks.
We run experiments on the Wilt Data Set with size (n) from 100 to 800. However, due to memory constraints, the amount of data we use in the FA task is 100 to 700. In each task, we set the power iteration method to run 50 times, which is enough to compute accurate results.

\subsubsection{Comparison of Resources Consumed by Traditional PCA SMPC Task and Decomposed PCA SMPC Task}
\begin{figure}[h]
	\centering
	\includegraphics[width=\linewidth]{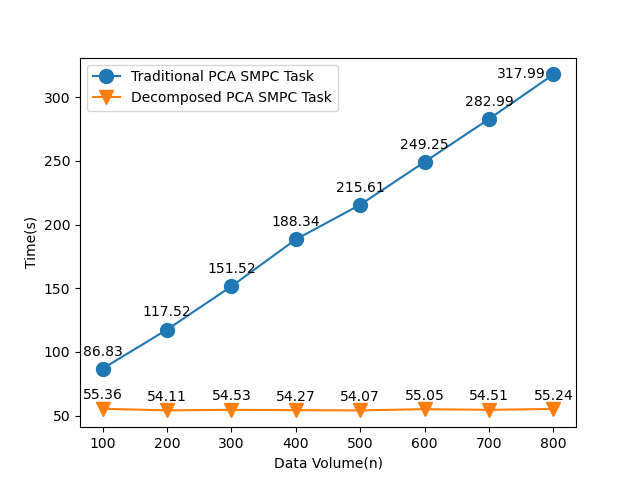}
	\caption{Time consumption of traditional PCA SMPC task and decomposed PCA SMPC task}
	\label{figure1}
\end{figure}

Figure \ref{figure1} shows the time consumption of two PCA SMPC tasks. It can be seen from Figure \ref{figure1} that for every increase of 100 data volumes, the traditional PCA SMPC task takes about 33 seconds longer. However, the running time of the decomposed PCA SMPC task has been stable at around 54 seconds.

\begin{figure}[h]	
	\centering
	\includegraphics[width=\linewidth]{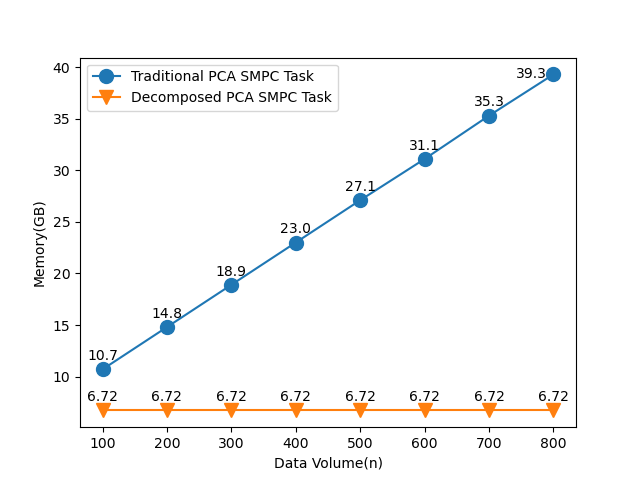}
	\caption{Memory consumption of traditional PCA SMPC task and decomposed PCA SMPC task}
	\label{figure2}
\end{figure}

Figure \ref{figure2} shows the memory consumption of both PCA tasks. It can be seen from Figure \ref{figure2} that for every increase of 100 data volumes, the memory consumed by the traditional PCA SMPC task increases by 4GB. However, the memory consumed by the decomposed PCA SMPC task has been stable at 6.72GB.

\begin{figure}[h]
	\centering
	\includegraphics[width=\linewidth]{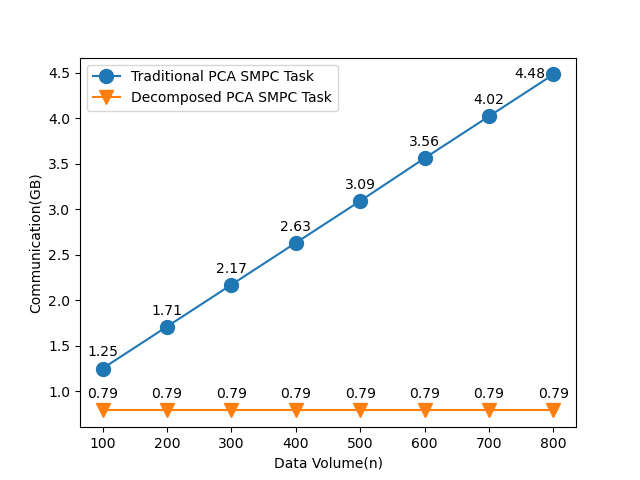}
	\caption{Communication consumption of traditional PCA SMPC task and decomposed PCA SMPC task}
	\label{figure3}
\end{figure}

Figure \ref{figure3} shows the communication consumption of both PCA tasks. It can be seen from Figure \ref{figure3} that for every increase of 100 data volumes, the communication consumed by the traditional PCA SMPC task increases by 0.46GB. However, the communication consumed by the decomposed PCA SMPC task has been stable at 0.79GB.


\subsubsection{Comparison of Resources Consumed by Traditional SVD SMPC Task and Decomposed SVD SMPC Task}
\begin{figure}[h]
	\centering
	\includegraphics[width=\linewidth]{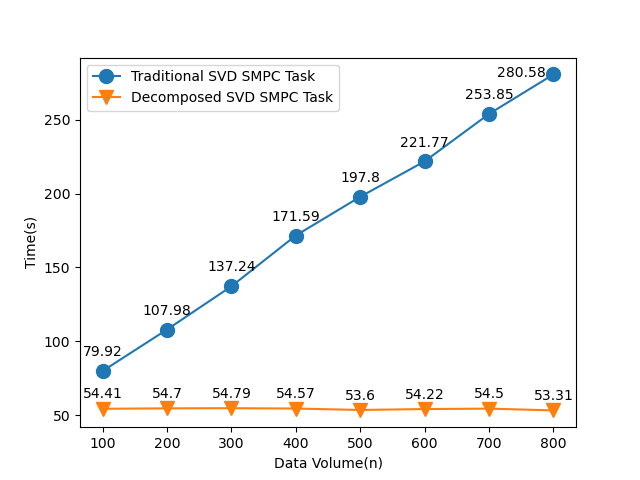}
	\caption{Time consumption of traditional SVD SMPC task and decomposed SVD SMPC task}
	\label{figure4}
\end{figure}

Figure \ref{figure4} shows the time consumption of both SVD SMPC tasks. It can be seen from Figure \ref{figure4} that for every increase of 100 data volumes, the traditional SVD SMPC task takes about 28.67 seconds longer. However, the running time of the decomposed SVD SMPC task has been stable at around 54 seconds.

\begin{figure}[h]
	\centering
	\includegraphics[width=\linewidth]{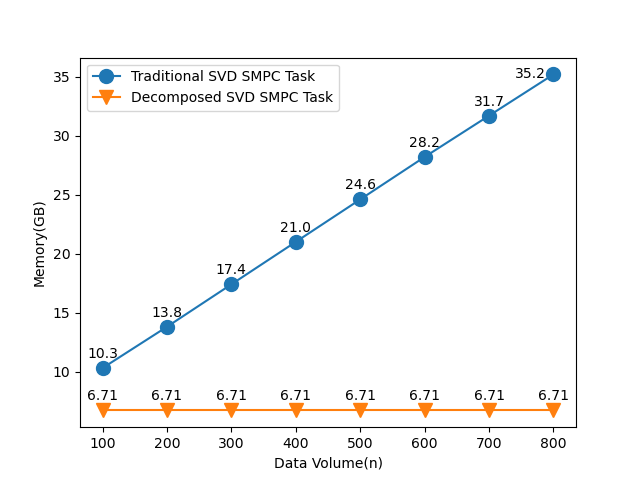}
	\caption{Memory consumption of traditional SVD SMPC task and decomposed SVD SMPC task}
	\label{figure5}
\end{figure}

Figure \ref{figure5} shows the memory consumption of both SVD tasks. It can be seen from Figure \ref{figure5} that for every increase of 100 data volumes, the memory consumed by the traditional SVD SMPC task increases by 3.6GB. However, the memory consumed by the decomposed SVD SMPC task has been stable at 6.71GB.

\begin{figure}[h]
	\centering
	\includegraphics[width=\linewidth]{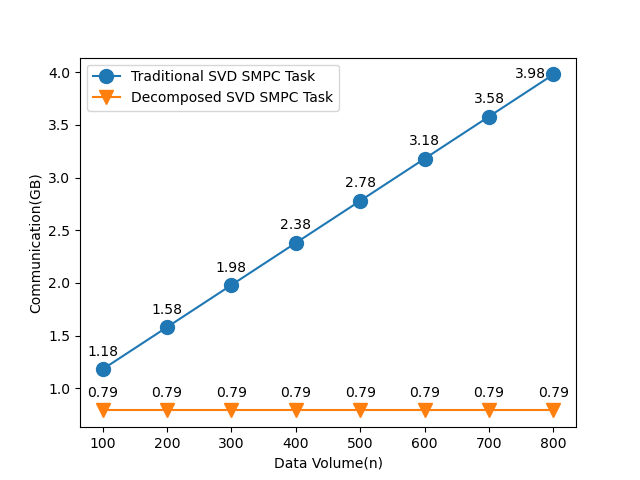}
	\caption{Communication consumption of traditional SVD SMPC task and decomposed SVD SMPC task}
	\label{figure6}
\end{figure}

Figure \ref{figure6} shows the communication consumption of both SVD tasks. It can be seen from Figure \ref{figure6} that for every increase of 100 data volumes, the communication consumed by the traditional SVD SMPC task increases by 0.40GB. However, the communication consumed by the decomposed SVD SMPC task has been stable at 0.79GB.
\subsubsection{Comparison of Resources Consumed by Traditional FA SMPC Task and Decomposed FA SMPC Task}
\begin{figure}[h]
	\centering
	\includegraphics[width=\linewidth]{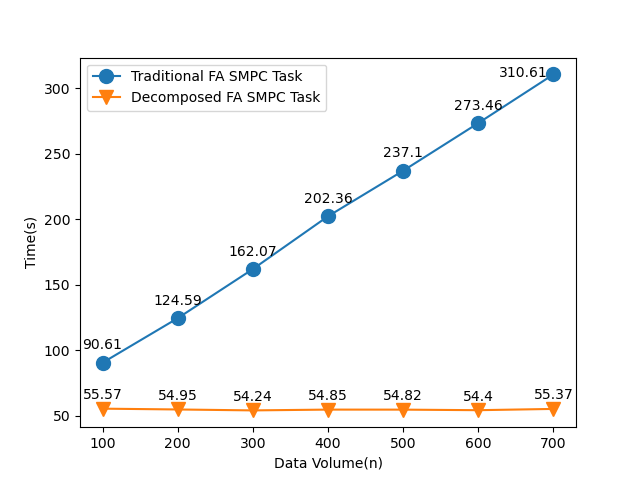}
	\caption{Time consumption of traditional FA SMPC task and decomposed FA SMPC task}
	\label{figure7}
\end{figure}

Figure \ref{figure7} shows the time consumption of both FA SMPC tasks. It can be seen from Figure \ref{figure7} that for every increase of 100 data volumes, the traditional FA SMPC task takes about 36.67 seconds longer. However, the running time of the decomposed FA SMPC task has been stable at around 54 seconds.

\begin{figure}[h]
	\centering
	\includegraphics[width=\linewidth]{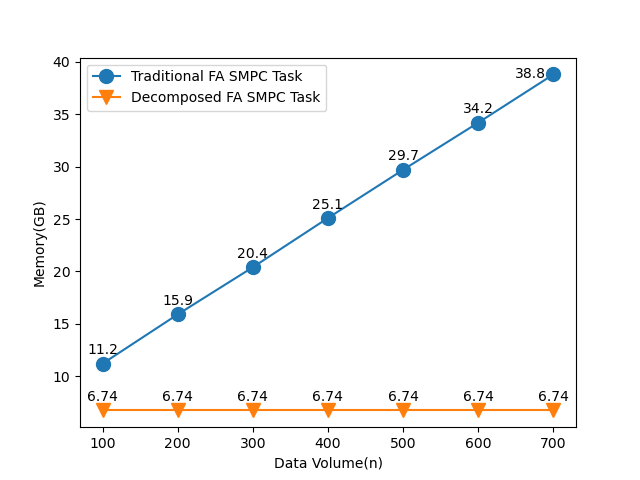}
	\caption{Memory consumption of traditional FA SMPC task and decomposed FA SMPC task}
	\label{figure8}
\end{figure}

Figure \ref{figure8} shows the memory consumption of both FA tasks. It can be seen from Figure \ref{figure8} that for every increase of 100 data volumes, the memory consumed by the traditional FA SMPC task increases by 4.6GB. However, the memory consumed by the decomposed FA SMPC task has been stable at 6.74GB.

\begin{figure}[h]
	\centering
	\includegraphics[width=\linewidth]{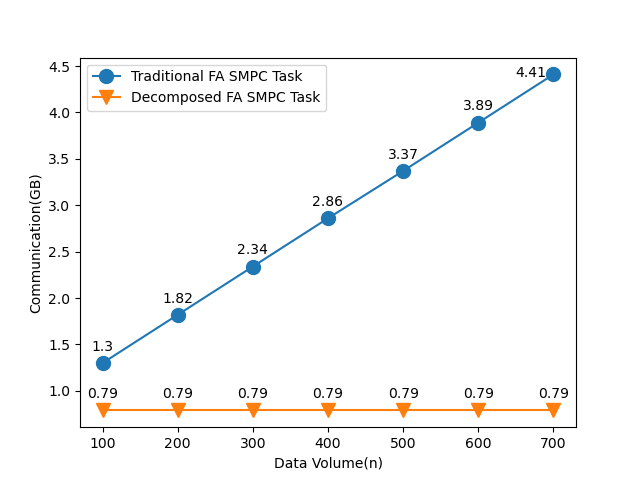}
	\caption{Communication consumption of traditional FA SMPC task and decomposed FA SMPC task}
	\label{figure9}
\end{figure}

Figure \ref{figure9} shows the communication consumption of both FA tasks. It can be seen from Figure \ref{figure9} that for every increase of 100 data volumes, the communication consumed by the traditional FA SMPC task increases by 0.52GB. However, the communication consumed by the decomposed FA SMPC task has been stable at 0.79GB.
\subsection{Experiments with More Data}
\label{section5.2}
In this section, we increase the amount of data used by the three decomposed SMPC tasks. We run experiments on the Accelerometer Data Set with size (n) from 7,650 to 76,500. In PCA and SVD algorithms, we set the power iteration method run 50 times, and the experimental results are almost identical to those calculated by the Python standard library. However, in the FA algorithm, we need to calculate the eigenvalues of the correlation coefficient matrix, which are very close, so we use the shift power iteration method and set it to run 300 times to get accurate results.

\subsubsection{Experimental Results of Decomposed PCA and SVD SMPC Tasks}
\begin{figure}[h]
	\centering
	\includegraphics[width=\linewidth]{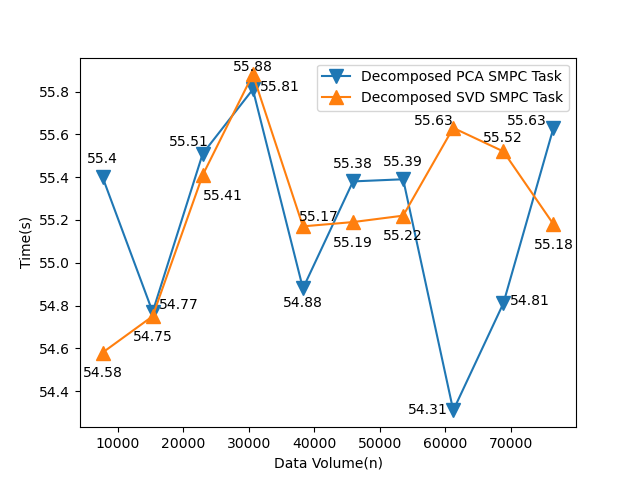}
	\caption{Time consumption of decomposed PCA and SVD SMPC tasks}
	\label{figure10}
\end{figure}

In Figure \ref{figure10}, the time consumption of the two SMPC tasks stabilized between 54 and 56 seconds, even when the data size is increased to 70,000.

\begin{figure}[h]
	\centering
	\includegraphics[width=\linewidth]{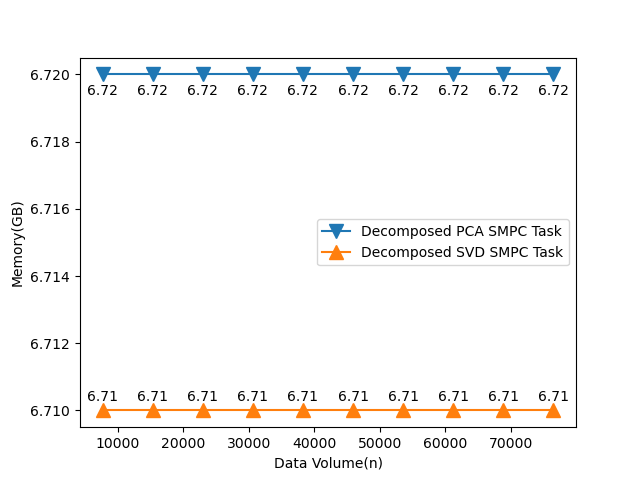}
	\caption{Memory consumption of decomposed PCA and SVD SMPC tasks}
	\label{figure11}
\end{figure}

In Figure \ref{figure11}, the memory consumption of the two SMPC tasks stabilized between 6.71GB and 6.72GB, even when the data size is increased to 70,000.

\begin{figure}[h]
	\centering
	\includegraphics[width=\linewidth]{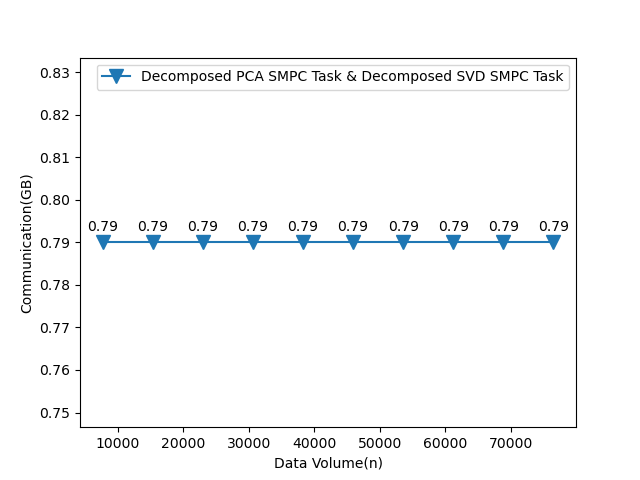}
	\caption{Communication consumption of decomposed PCA and SVD SMPC tasks}
	\label{figure12}
\end{figure}

In Figure \ref{figure12}, the communication consumption of the two SMPC tasks stabilized at 0.79GB, even when the data size is increased to 70,000.
\subsubsection{Experimental Results of Decomposed FA SMPC Task}
\begin{figure}[h]
	\centering
	\includegraphics[width=\linewidth]{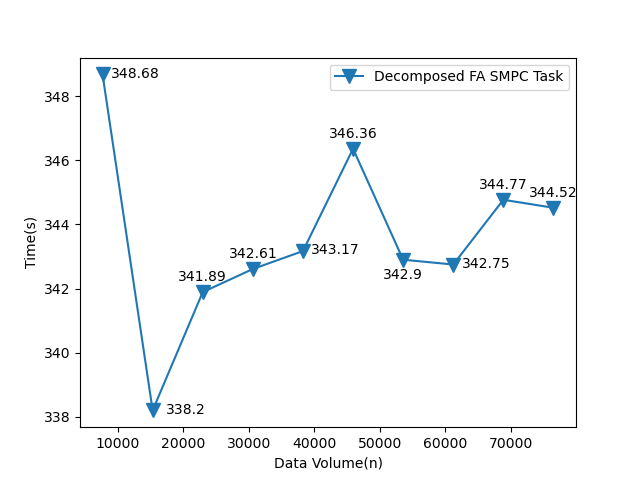}
	\caption{Time consumption of decomposed FA SMPC task}
	\label{figure13}
\end{figure}

In Figure \ref{figure13}, the time consumption of the decomposed FA SMPC task is stable between 338 and 349 seconds even when the data size is increased to 70,000.

\begin{figure}[h]
	\centering
	\includegraphics[width=\linewidth]{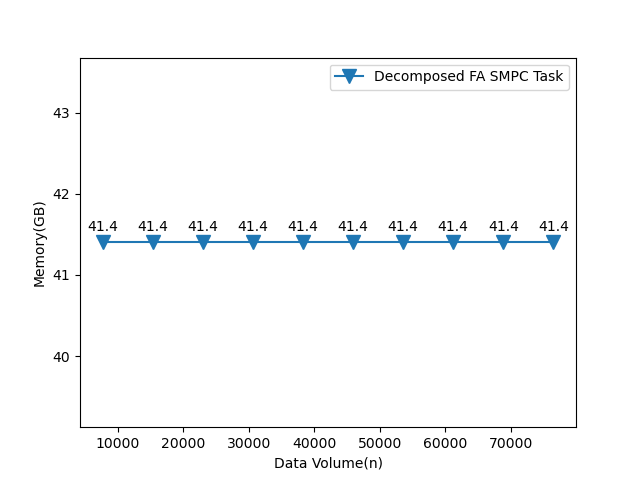}
	\caption{Memory consumption of decomposed FA SMPC task}
	\label{figure14}
\end{figure}

In Figure \ref{figure14}, the memory consumption of the decomposed FA SMPC task is stable at 41.4GB even when the data size is increased to 70,000.

\begin{figure}[h]
	\centering
	\includegraphics[width=\linewidth]{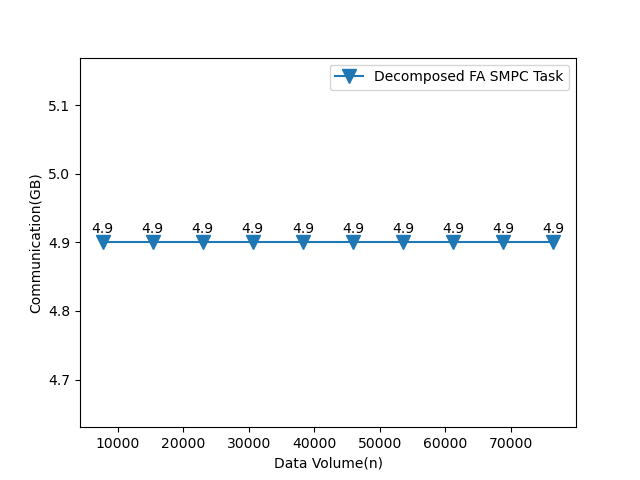}
	\caption{Communication consumption of decomposed FA SMPC task}
	\label{figure15}
\end{figure}

In Figure \ref{figure15}, the communication consumption of the decomposed FA SMPC task is stable at 4.9GB even when the data size is increased to 70,000.

\subsection{Accuracy of Experimental Results}
\label{section5.3}
Through the decomposed PCA SMPC task, we obtain the mean of all data, the eigenvalues corresponding to the covariance matrix of the data, and the eigenvector matrix corresponding to the eigenvalues. Through the decomposed SVD SMPC task, we obtain the singular value of the data and the right singular matrix corresponding to the singular value. Through the decomposed FA SMPC task, we obtain the principal factors and factor load matrices of the data. 

Then we retrain on the total dataset using the Python standard library, such as NumPy \cite{harris2020array}, sklearn \cite{scikit-learn} and factor\_analyzer \cite{factor_analyzer}. The results are almost identical to the results obtained by the three SMPC tasks. Our Python code is also in our project.

\subsection{Analysis of Experimental Results}
\label{section5.4}
From our experiments in Section \ref{section5.1} and Section \ref{section5.2}, we can see that after decomposing a SMPC task using our theory, the computing time, memory and communication consumption of the task are greatly reduced, and do not increase linearly with the increase of data volume. The main reason is that we separate a large number of calculations into local tasks, and no longer need to generate a circuit gate for each raw data, so the overall circuit size is greatly reduced. In addition, we only build circuits when combining data, so the number of circuits is also deterministic. Thus the consumption of resources has been stable in a range.

\section{Conclusion}

We propose a novel theory in a semi-honest adversary model to address the problem that the resources consumed by SMPC tasks increase substantially with the amount of data. Then we analyze the security of our theory and prove that it is safe. We decompose three linear dimension reduction SMPC tasks. Then we compare the resources consumption of the three ML SMPC tasks before and after decomposition. The experimental results show that our theory is very efficient. The resources consumed by the three ML SMPC tasks before decomposition increase linearly with the increase of the amount of data, while the resources consumed by the three ML SMPC tasks after decomposition are always stable within a certain range.

\section*{Acknowledgment}
This work is supported by the National Key R\&D Program of China under Grand No.2021YFB2012202, the Key
Research \& Development Plan of Hubei Province of China
under Grant No.2021BAA171, No.2021BAA038, and the
project of Science, Technology and Innovation Commission of Shenzhen Municipality of China under Grant No.
JCYJ20210324120002006 and JSGG20210802153009028.

\bibliographystyle{IEEEtran}
\bibliography{IEEEabrv,ieee}

\end{document}